\documentclass[pra,aps,showpacs,twocolumn]{revtex4-1}
\usepackage{amsmath}
\usepackage{amssymb}
\usepackage{graphicx}
\usepackage{epstopdf}
\usepackage{mathtools}
\usepackage{appendix}
\usepackage[bookmarks=false]{hyperref}
\hypersetup{colorlinks=true,citecolor=blue,linkcolor=blue,urlcolor=blue,pdfstartview=FitH,bookmarksopen=true}
\usepackage{xcolor}
\usepackage{soul}
\usepackage{times,txfonts}
\setstcolor{blue}
\setulcolor{red}

\newcommand{\ket}[1]{|#1\rangle}

\begin{document}

\title{Nonadiabatic holonomic multiqubit controlled gates}
\author{P. Z. Zhao}
\affiliation{Department of Physics, Shandong University, Jinan 250100, China}
\author{G. F. Xu}
\email{sduxgf@163.com}
\affiliation{Department of Physics, Shandong University, Jinan 250100, China}
\author{D. M. Tong}
\email{tdm@sdu.edu.cn}
\affiliation{Department of Physics, Shandong University, Jinan 250100, China}
\date{\today}

\begin{abstract}
Previous schemes of nonadiabatic holonomic quantum computation were focused mainly on realizing a universal set of elementary gates. Multiqubit controlled gates could be built by decomposing them into a series of the universal gates. In this article, we propose an approach for realizing nonadiabatic holonomic multiqubit controlled gates in which a $(n+1)$-qubit controlled-$(\boldsymbol{\mathrm{n}\cdot \mathrm{\sigma}})$ gate is realized by $(2n-1)$ basic operations instead of decomposing it into the universal gates, whereas an $(n+1)$-qubit controlled arbitrary rotation gate can be obtained by combining only two such controlled-$(\boldsymbol{\mathrm{n}\cdot \mathrm{\sigma}})$ gates. Our scheme greatly reduces the operations of nonadiabatic holonomic quantum computation.
\end{abstract}

\maketitle
Quantum computation is founded on quantum-mechanical principles, and is performed by unitary quantum gates. This creates its superiorities that are not available for classical computation. However, such superiorities rely on the ability to perform high-fidelity quantum gates. Two main challenges in achieving such high-fidelity gates are to reduce control errors of a quantum system and to avoid decoherence caused by the environment. To overcome these problems, various proposals of quantum computation with noise-resilience features have been proposed.  A promising such proposal is nonadiabatic holonomic quantum computation.

Nonadiabatic holonomic quantum computation \cite{Sjoqvist,Xu} is based on nonadiabatic non-Abelian geometric phases \cite{Anandan}. It is realized by using a quantum system with a subspace satisfying both the cyclic evolution and the parallel transport conditions. Consider a $N-$dimensional quantum system defined by $H(t)$, of which the evolution operator reads $U(t,0) ={\bf T} \exp{i\int_0^tH(t')dt'}$. If there exists a time-dependent $L-$dimensional subspace $\mathcal{S}(t)$ spanned by the orthonormal vectors $\{\ket{\phi_k(t)}=U(t,0)\ket{\phi_k(0)} \}_{k=1}^L$  that satisfy the two conditions: $ \textrm{(i)}$  $\sum_{k=1}^L |\phi_{k}(\tau)\rangle \langle
\phi_{k}(\tau)|=\sum_{k=1}^L |\phi_{k}(0)\rangle \langle \phi_{k}(0)|$ with $\tau$ being the evolution period,  and $\textrm{(ii)}$  $\langle\phi_{k}(t)|H(t)|\phi_{l}(t)\rangle=0,\ k,l=1,...,L,$  then the unitary transformation $U(\tau,0)$ is a holonomic gate on the $L-$dimensional subspace $\mathcal{S}(0)$ spanned by $\{ \ket{\phi_k(0)} \}_{k=1}^L$. This gate is only dependent on evolution paths but independent of evolution details, being robust against control errors.

Nonadiabatic holonomic quantum computation is a gradual development of the early adiabatic geometric quantum computation \cite{Jones} based on Berry phases \cite{Berry}, adiabatic holonomic quantum computation \cite{Zanardi,Duan} based on adiabatic non-Abelian geometric phases \cite{Wilczek}, and nonadiabatic geometric quantum computation \cite{WangXB,Zhu1} based on  Aharonov-Anandan phases  \cite{Aharonov}.
It shares all the holonomic nature of its adiabatic counterpart whereas avoiding the long run-time requirement.
Due to the merits of both its robustness against control errors and its rapidity without the speed limit of the adiabatic evolution, nonadiabatic holonomic quantum computation has received increasing attention. Since its original proposal \cite{Sjoqvist,Xu},
many schemes of its implementation have been put forward based on various physical systems
\cite{Johansson2012,Spiegelberg2013,Liang,Zhang,Mousolou2014,ZhangT2,Xu2015,Xue,E2016,You,S2016,Sun,Xue2016,
Xue2017,Zhao,Zhao2017,Su2017,Xu2017,Xu2017PRA,Zhao2018,Mousolou2017,Xue2018,Zhang2018,XuGF2018}. Encouragingly, nonadiabatic holonomic quantum computation has been experimentally demonstrated with nuclear magnetic resonance \cite{Long,Long2017}, superconducting circuits \cite{Abdumalikov,Xu2018,Danilin,Egger,Yan,Yin}, and nitrogen-vacancy centers in diamond \cite{Arroyo,Duan2014,Sekiguchi,Zhou,Nagata,Ishida}.

These previous schemes of nonadiabatic holonomic quantum computation were focused mainly on realizing a universal set of elementary gates, e.g., two noncommuting one-qubit gates and a two-qubit entangling gate \cite{Huangnew}. A multiqubit gate could be built by decomposing it into a series of universal gates. Particularly, a $(n+1)$-qubit controlled gate, as one important family of quantum gates being widely used in quantum information processing \cite{Shor,Steane,Grover,Vandersypen,Joshi,Yang,Ota}, needs to be decomposed into $(2^{n+1}-3)$ two-qubit controlled gates \cite{Barenco,Goto}. The decomposition becomes complicated as the number of qubits increases.
It makes us attempt to find a new approach by which nonadiabatic holonomic multiqubit controlled gates can be realized with fewer operations, i.e., they can be more effectively performed without decomposing them into so many elementary gates. This is an interesting topic,
as fewer operations imply less accumulation of control errors and less exposure time to decoherence, which results in higher-fidelity gates.

In this article, we propose a scheme to realize nonadiabatic holonomic multiqubit controlled gates. Our physical model is a set of ions trapped in a linear trap. Each ion has three levels corresponding to states $|0\rangle$, $|1\rangle$ and $|e\rangle$, which form a $\Lambda$ configuration. The two lower states $|0\rangle$ and $|1\rangle$ play the role of a qubit, whereas the excited-state $|e\rangle$ acts as an auxiliary.  The computational space for the $(n+1)$-qubit gate, denoted as $\mathcal{S}_{n+1}$, is spanned by $\{|b_1b_2\cdots b_{n+1}\rangle,~b_1,b_2,\cdots,b_{n+1}=0,1\}$.
We will show how to realize a holonomic $(n+1)$-qubit controlled-($\boldsymbol{\mathrm{n}\cdot\sigma}$) gate, i.e. a controlled $\pi$-rotation gate, whereas a holonomic $(n+1)$-qubit controlled arbitrary rotation gate can be realized by combining two of these gates. Here, $\boldsymbol{\mathrm{n}} =(\sin\theta\cos\varphi,\sin\theta\sin\varphi,\cos\theta)$ represents the rotation axis, and $\boldsymbol{\sigma}$ is the Pauli operator. For simplicity, we use
$U_{{C^n-\boldsymbol{\mathrm{n}\cdot\sigma}}}$ to denote the gate, which can be explicitly expressed as
\begin{align}
U_{{C^n-\boldsymbol{\mathrm{n}\cdot\sigma}}}=\left[I^{\otimes n}-(|1\rangle\langle1|)^{\otimes n}\right]\otimes I+
(|1\rangle\langle1|)^{\otimes n}\otimes\boldsymbol{\mathrm{n}\cdot\sigma}
\end{align}
with $I$ being a $2\times 2$ identity matrix.
The main challenge of realizing this gate is how to achieve the multi-ion couplings that make the evolution of the qubits fulfill the two holonomic conditions $\textrm{(i)}$ and $\textrm{(ii)}$. By reducing the $(n+1)$-ion couplings into a series of experimentally achievable two-ion couplings \cite{SM1999,SM2000,Benhelm,Webb,Shapira} and designing the Hamiltonian of the two-ion pairs, we resolve the challenge and realize the controlled operations on the target qubit.

{\it First, we show how to realize a nonadiabatic holonomic two-qubit controlled-($\boldsymbol{\mathrm{n}\cdot\sigma}$) gate,
 \begin{align}
U_{{C^1-\boldsymbol{\mathrm{n}\cdot\sigma}}}=|0\rangle\langle0|\otimes I+|1\rangle\langle1|\otimes\boldsymbol{\mathrm{n}\cdot\sigma}.
\label{eq}
\end{align}}

\begin{figure*}[t]
  \includegraphics[scale=0.25]{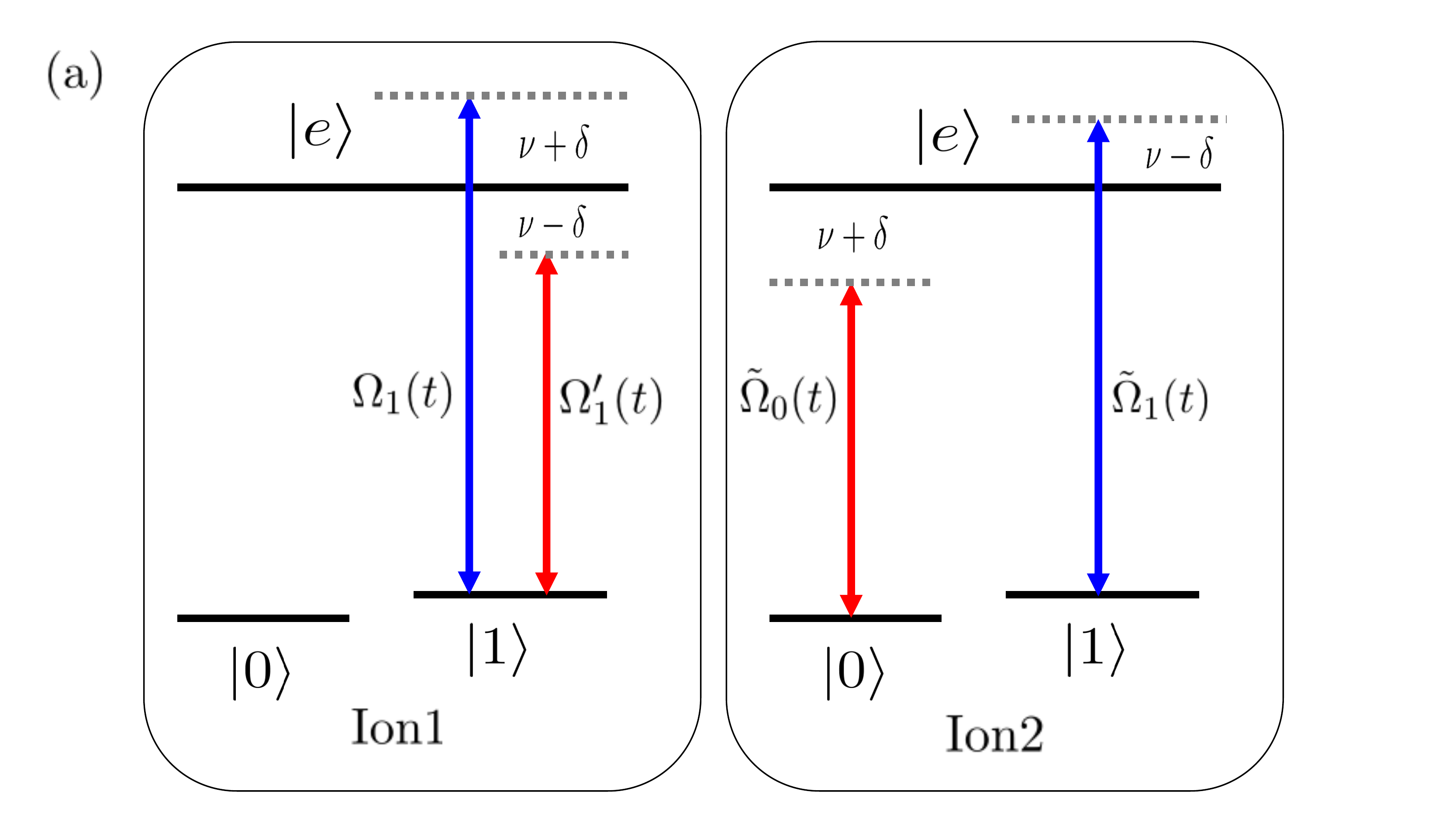}
  \includegraphics[scale=0.25]{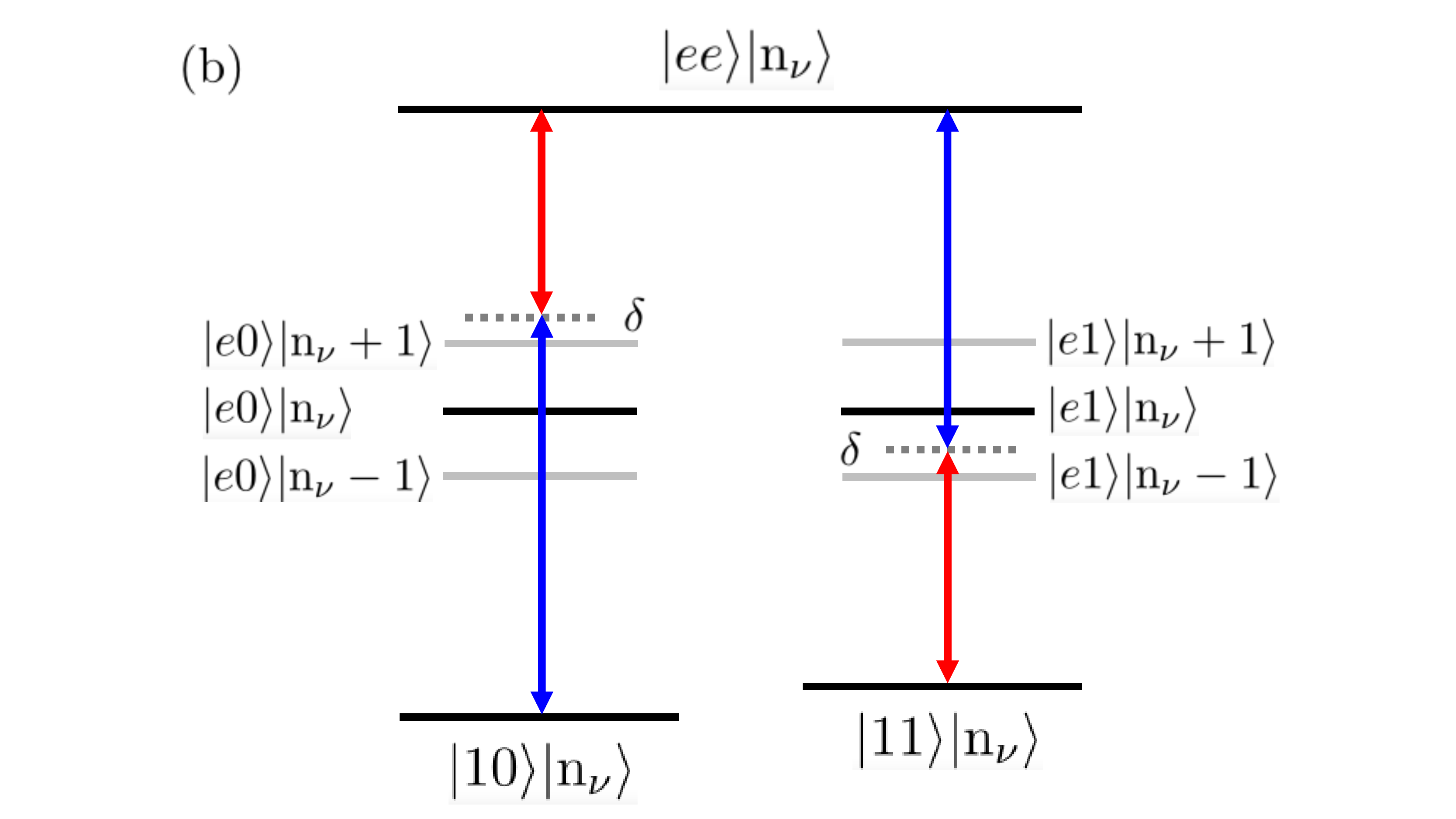}
  \caption{(Color online) Schematic for ion-laser interaction.
   (a) Level configuration of two three-level ions driven by lasers, corresponding to the Hamiltonian in Eq. (\ref{eq1}).
   (b) Level diagram for collective transition of the two ions, corresponding to the effective Hamiltonian in Eq. (\ref{eq2}). The only resonant transitions are $|10\rangle|n_\nu\rangle\leftrightarrow|ee\rangle|n_\nu\rangle$ and $|11\rangle|n_\nu\rangle\leftrightarrow|ee\rangle|n_\nu\rangle$ with the aid of the intermediated states $|e0\rangle|n_\nu+1\rangle$ and $|e1\rangle|n_\nu-1\rangle$.}
   \label{Fig1}
\end{figure*}

Consider two ions trapped in a linear trap.  A pair of bichromatic lasers are applied to each ion to drive the transitions $|0\rangle\leftrightarrow|e\rangle$ or/and  $|1\rangle\leftrightarrow|e\rangle$, as shown in Fig. \ref{Fig1}(a).
For ion $1$, we use a blue sideband laser with detuning  $-(\nu+\delta)$ and Rabi frequency $\Omega_{1}(t)$ and a red sideband laser with detuning $(\nu-\delta)$ and Rabi frequency $\Omega^{\prime}_{1}(t)$ to drive the transition $|1\rangle\leftrightarrow|e\rangle$.
For ion $2$, we use a red sideband laser with  detuning $\nu+\delta$ and Rabi frequency $\tilde{\Omega}_{0}(t)$ to drive the transition $|0\rangle\leftrightarrow|e\rangle$, and a blue sideband laser with detuning $-(\nu-\delta)$ and Rabi frequency $\tilde{\Omega}_{1}(t)$ to drive the transition $|1\rangle\leftrightarrow|e\rangle$.  Here, $\nu$ is the frequency of the vibrational mode for trapped ions, and $\delta$ is an additional detuning.  In the rotating frame and with the rotating-wave approximation, the Hamiltonian of the two-ion system reads
\begin{align}
H(t)=&i\eta\Omega_{1}(t)e^{-i\delta t}a^{\dag}|e\rangle_{11}\langle1|+i\eta\Omega^{\prime}_{1}(t)e^{-i\delta t}a|e\rangle_{11}\langle1| \notag\\
&+i\eta\tilde{\Omega}_{0}(t)e^{i\delta t}a|e\rangle_{22}\langle0|+i\eta\tilde{\Omega}_{1}(t)e^{
i\delta t}a^{\dag}|e\rangle_{22}\langle1|+\mathrm{H.c.} \label{eq1}
\end{align}
in the Lamb-Dicke regime. Here, $|m\rangle_{j}$ $(m=0,1,e)$ represents state $|m\rangle$ of the $j$th ion, $a$ and ${a}^\dagger$ are the annihilation and creation operators of the vibrational mode, and $\eta$ is the Lamb$-$Dicke parameter that satisfies $\eta^{2}(n_\nu+1)\ll1$ with $n_\nu$ being the quantum number of the vibrational mode.

If the large detuning condition $\delta\gg\eta\Omega_{1}(t),$ $\eta\Omega^{\prime}_{1}(t),$ $\eta\tilde{\Omega}_{0}(t),$ $\eta\tilde{\Omega}_{1}(t)$ is satisfied, the single-ion transitions $|0\rangle\leftrightarrow|e\rangle$ and $|1\rangle\leftrightarrow|e\rangle$ are strongly suppressed whereas only the double-ion transitions $|10\rangle\leftrightarrow|ee\rangle$ and $|11\rangle\leftrightarrow|ee\rangle$ are allowed due to vibrational energy exchanging between two ions, as shown in Fig. \ref{Fig1}(b). In this case, the Hamiltonian in Eq. (\ref{eq1}) is reduced to an effective one \cite{Tong1},
\begin{align}
H_{\mathrm{eff}}(t)=\Omega_{10}(t)|ee\rangle\langle10|+\Omega_{11}(t)|ee\rangle\langle11|+\mathrm{H.c}, \label{eq2}
\end{align}
where $\Omega_{10}(t)=-\eta^{2}\Omega_{1}(t)\tilde{\Omega}_{0}(t)/\delta$, $\Omega_{11}(t)=\eta^{2}\Omega^{\prime}_{1}(t)\tilde{\Omega}_{1}(t)/\delta$.

To obtain the two-qubit controlled-($\boldsymbol{\mathrm{n}\cdot\sigma}$) gate with $\boldsymbol{\mathrm{n}} =(\sin\theta\cos\varphi,\sin\theta\sin\varphi,\cos\theta)$, we set $\Omega_{10}(t)=\Omega(t)\sin(\theta/2)e^{i\varphi}$ and $\Omega_{11}(t)=-\Omega(t)\cos(\theta/2)$, where $\Omega(t)$ is a time dependent parameter. In this case, the effective Hamiltonians at different times commute with each other. We then have the evolution operator,
\begin{align}
U(t,0)=&\exp\left[-i\int^{t}_{0}H_{\mathrm{eff}}(t^{\prime})dt^{\prime}\right]
\notag\\
=&|00\rangle\langle00|+|01\rangle\langle01|+|\mathrm{D_2}\rangle\langle\mathrm{D_2}| \notag\\
&
+\cos\phi_{t}\left(|\mathrm{B_2}\rangle\langle\mathrm{B_2}|+|ee\rangle\langle ee|\right)\notag\\
&
-i\sin\phi_{t}\left(|\mathrm{B_2}\rangle\langle ee|+|ee\rangle\langle\mathrm{B_2}|\right)
\notag\\
&+|e0\rangle\langle e0|+|e1\rangle\langle e1|+|0e\rangle\langle0e|+|1e\rangle\langle1e|
 \label{eq3}
\end{align}
with  $|\mathrm{D_2}\rangle=\cos(\theta/2)|10\rangle+\sin(\theta/2)e^{i\varphi}|11\rangle$,  $|\mathrm{B_2}\rangle=\sin(\theta/2)e^{-i\varphi}|10\rangle-\cos(\theta/2)|11\rangle$ and $\phi_{t}=\int^{t}_{0}\Omega(t^{\prime})dt^{\prime}$.
If the evolution period $\tau$ is taken to satisfy
\begin{align}
\phi_{\tau}=\int^{\tau}_{0}\Omega(t)dt=\pi,
\end{align}
there is
\begin{align}
U(\tau,0)=&|00\rangle\langle00|+|01\rangle\langle01|+|\mathrm{D_2}\rangle\langle\mathrm{D_2}|-
|\mathrm{B_2}\rangle\langle\mathrm{B_2}|-|ee\rangle\langle ee|
\notag\\
&+|e0\rangle\langle e0|+|e1\rangle\langle e1|+|0e\rangle\langle0e|+|1e\rangle\langle1e|. \label{eq4}
\end{align}
It shows that the role of the evolution operator $U(\tau)$ on the computational subspace is equivalent to
\begin{align}
U_{{C^1-\boldsymbol{\mathrm{n}\cdot\sigma}}}=&|00\rangle\langle00|+|01\rangle\langle01|+|\mathrm{D_2}\rangle\langle\mathrm{D_2}|-
|\mathrm{B_2}\rangle\langle\mathrm{B_2}|,
\label{uc1}
\end{align}
which can be recast as Eq. (\ref{eq}) since $|00\rangle\langle00|+|01\rangle\langle01|=|0\rangle\langle0|\otimes I$ and $|\mathrm{D_2}\rangle\langle\mathrm{D_2}|-
|\mathrm{B_2}\rangle\langle\mathrm{B_2}|=|1\rangle\langle1|\otimes\boldsymbol{\mathrm{n}\cdot\sigma}$.

We now demonstrate that $U_{{C^1-\boldsymbol{\mathrm{n}\cdot\sigma}}}$ is a holonomic gate, i.e., the two conditions  $(\mathrm{i})$ and
$(\mathrm{ii})$ are fulfilled. Equation (\ref{eq3}) clearly shows  that a state initially residing in the computational space $\mathcal{S}_2$ may evolve into the outside of the subspace during $t\in(0,\tau)$ but returns back to it at $t=\tau$, i.e.,
condition $(\mathrm{i})$ is satisfied. Furthermore, with the aid of the relation $[H_{\mathrm{eff}}(t),U(t,0)]=0$, it is easy to verify that $\langle\mu(t)|H_{\mathrm{eff}}(t)|\nu(t)\rangle=\langle\mu|H_{\mathrm{eff}}(t)|\nu\rangle=0$, where $|\mu(t)\rangle=U(t,0)|\mu\rangle$, $|\nu(t)\rangle=U(t,0)|\nu\rangle$, and $|\mu\rangle,|\nu\rangle\in \mathcal{S}_2$, i.e., condition $(\mathrm{ii})$ is satisfied too. Therefore, the unitary operator $U(\tau,0)$ plays a holonomic gate on the computational subspace.

{\it Second, we show how to realize a nonadiabatic holonomic three-qubit controlled-($\boldsymbol{\mathrm{n}\cdot\sigma}$)} gate,
\begin{align}
U_{{C^2-\boldsymbol{\mathrm{n}\cdot\sigma}}}=\left[I^{\otimes 2}-(|1\rangle\langle1|)^{\otimes 2}\right]\otimes I+
(|1\rangle\langle1|)^{\otimes 2}\otimes\boldsymbol{\mathrm{n}\cdot\sigma}.
\label{eq5}
\end{align}

For the three-qubit controlled gate, we consider three ions trapped in a linear trap. We need to use a piecewise time-dependent Hamiltonian, i.e., we divide the whole evolution time $\tau$ into three intervals: $0\le t \le \tau_1$, $\tau_1\le t \le \tau_2$ and $\tau_2\le t \le \tau$, in each of which a special Hamiltonian is chosen.

In the first interval, a red sideband laser with detuning $(\nu-\delta)$ and Rabi frequency $\Omega^\prime_1(t)$ is applied to ion $1$ to drive the transition $|1\rangle\leftrightarrow|e\rangle$, and  a blue sideband laser with detuning $-(\nu-\delta)$ and Rabi frequency $\tilde{\Omega}_1(t)$ is applied to ion $2$ to drive the transition $|1\rangle\leftrightarrow|e\rangle$. In this case, the Hamiltonian can be written as   $H(t)=i\eta\Omega^{\prime}_{1}(t)e^{-i\delta t}a|e\rangle_{11}\langle1| +i\eta\tilde{\Omega}_{1}(t)e^{
i\delta t}a^{\dag}|e\rangle_{22}\langle1|+\mathrm{H.c.}$, which is a special case of Eq. (\ref{eq1}) with $\Omega_{1}(t)=\tilde{\Omega}_{0}(t)=0$, and therefore the effective  Hamiltonian has the same form as Eq. (\ref{eq2}) but with $\Omega_{10}(t)=-\eta^{2}\Omega_{1}(t)\tilde{\Omega}_{0}(t)/\delta=0$, i.e., the effective Hamiltonian used during $t\in[0,\tau_1]$ is
\begin{align}
H_{\mathrm{eff}}^{(1)}(t)=\Omega_{11}(t)|ee\rangle\langle11|+\mathrm{H.c.},\label{T1}
\end{align}
which acts on ions $1$ and $2$.

In the second interval, two lasers, one of which is with detuning  $-(\nu+\delta)$ and Rabi frequency $\Omega_{1}(t)$, and another is with $(\nu-\delta)$ and $\Omega^{\prime}_{1}(t)$,  are applied to ion $2$ to drive the transition $|1\rangle\leftrightarrow|e\rangle$, and two lasers, one of which is with detuning  $-(\nu+\delta)$ and Rabi frequency $\tilde{\Omega}_0(t)$, and another is with $(\nu-\delta)$ and $\tilde{\Omega}^{\prime}_{1}(t)$, are applied to ion $3$  to drive the transitions $|0\rangle\leftrightarrow|e\rangle$ and $|1\rangle\leftrightarrow|e\rangle$, respectively. The Hamiltonian reads $H(t)=i\eta\Omega_{1}(t)e^{-i\delta t}a^{\dag}|e\rangle_{22}\langle1|
+i\eta\Omega^{\prime}_{1}(t)e^{-i\delta t}a|e\rangle_{22}\langle1|+
i\eta\tilde{\Omega}_{0}(t)e^{-i\delta t}a^{\dag}|e\rangle_{33}\langle0|
+i\eta\tilde{\Omega}_{1}(t)e^{-i\delta t}a|e\rangle_{33}\langle1|+\mathrm{H.c.}$
Under the large detuning condition $\delta\gg\eta\Omega_{1}(t),$ $\eta\Omega^{\prime}_{1}(t),$ $\eta\tilde{\Omega}_{0}(t),$ $\eta\tilde{\Omega}_{1}(t)$,
we can obtain the effective Hamiltonian used during $t\in[\tau_1,\tau_2]$ \cite{Tong1},
\begin{align}
H_{\mathrm{eff}}^{(2)}(t)=\Omega_{e0}(t)|1e\rangle\langle e0|+\Omega_{e1}(t)|1e\rangle\langle e1|+\mathrm{H.c.},\label{T2}
\end{align}
which acts on ions $2$ and $3$. Here, $\Omega_{e0}(t)=\eta^{2}\Omega^{*}_{1}(t)\tilde{\Omega}_{0}(t)/\delta$,
$\Omega_{e1}(t)=-\eta^{2}[\Omega^{\prime}_{1}(t)]^{*}\tilde{\Omega}_{1}(t)/\delta$.

In the third interval, i.e., during $t\in[\tau_2,\tau]$, we use the same Hamiltonian as that in the first interval,
\begin{align}
H_{\mathrm{eff}}^{(3)}(t)=\Omega_{11}(t)|ee\rangle\langle11|+\mathrm{H.c.},
\end{align}
which acts on ions $1$ and $2$.

To obtain the nonadiabatic holonomic three-qubit controlled-$(\boldsymbol{\mathrm{n}\cdot\sigma})$ gate,
we choose $\Omega_{11}(t)$ to be real and set $\Omega_{e0}(t)=\Omega(t)\cos(\theta/2)$ and $\Omega_{e1}(t)=\Omega(t)\sin(\theta/2)e^{-i\varphi}$, where $\Omega(t)$ is a time dependent parameter. In this case, $[H_{\mathrm{eff}}^{(k)}(t),H_{\mathrm{eff}}^{(k)}(t^\prime)]=0$, $k=1,2,3$, and the evolution operator can be expressed as
\begin{align}
U(t,0)=\begin{dcases}
~\exp\left[-i\int^{t}_{0}H_{\mathrm{eff}}^{(1)}(t^{\prime})dt^{\prime}\right],&t\in[0,\tau_{1}], \\
~U(t,\tau_1)U(\tau_{1},0),&t\in[\tau_{1},\tau_{2}], \\
~U(t,\tau_2)U(\tau_{2},0),&t\in[\tau_{2},\tau],
\end{dcases}
\label{ut}
\end{align}
where $U(t,\tau_1)=\exp\left[-i\int^{t}_{\tau_{1}}H_{\mathrm{eff}}^{(2)}(t^{\prime})dt^{\prime}\right]$ and $U(t, \tau_2)=\exp\left[-i\int^{t}_{\tau_{2}}H_{\mathrm{eff}}^{(3)}(t^{\prime})dt^{\prime}\right]$.

We let $|\mathrm{D_3}\rangle=\cos(\theta/2)|110\rangle+\sin(\theta/2)e^{i\varphi}|111\rangle$ and  $|\mathrm{B_3}\rangle=\sin(\theta/2)e^{-i\varphi}|110\rangle-\cos(\theta/2)|111\rangle$, which span the same subspace as $\textrm{Span}\{|110\rangle,|111\rangle\}$.
If the time intervals $\tau_{1}$, $\tau_{2}$, and $\tau$ are taken to satisfy
\begin{align}
\int^{\tau_{1}}_{0}\Omega_{11}(t)dt=\frac{\pi}{2},
\int^{\tau_{2}}_{\tau_{1}}\Omega(t)dt=\pi,
\int^{\tau}_{\tau_{2}}\Omega_{11}(t)dt=\frac{\pi}{2},
\end{align}
we have
\begin{align}
U(\tau,0)|\mathrm{D_3}\rangle=|\mathrm{D_3}\rangle,~~U(\tau,0)|\mathrm{B_3}\rangle=-|\mathrm{B_3}\rangle,
\end{align}
and except $\{|D\rangle,|B\rangle\}$, i.e. $\{|110\rangle,|111\rangle\}$, all the other computational basis $\{|000\rangle,|001\rangle,|010\rangle,|011\rangle,|100\rangle,|101\rangle\}$  are unchanged in the whole evolution.
Consequently, if we focus only on the computational subspace $\mathcal{S}_3$, the role of $U(\tau,0)$ is equivalent to
\begin{align}
U_{{C^2-\boldsymbol{\mathrm{n}\cdot\sigma}}}=&\left(|00\rangle\langle00|+|01\rangle\langle01|+|10\rangle\langle10|\right)\otimes I
\notag\\
&+|\mathrm{D_3}\rangle\langle\mathrm{D_3}|-|\mathrm{B_3}\rangle\langle\mathrm{B_3}|,
\end{align}
which is Eq. (\ref{eq5}) since $|00\rangle\langle00|+|01\rangle\langle01|+|10\rangle\langle10|=I^{\otimes 2}-|11\rangle\langle 11|$ and $|\mathrm{D_3}\rangle\langle\mathrm{D_3}|-
|\mathrm{B_3}\rangle\langle\mathrm{B_3}|=|11\rangle\langle11|\otimes\boldsymbol{\mathrm{n}\cdot\sigma}$.

We now demonstrate that $U_{{C^2-\boldsymbol{\mathrm{n}\cdot\sigma}}}$ is a holonomic gate. The above discussion clearly shows  that
condition $(\mathrm{i})$ is satisfied. Furthermore, since $U(t,0)$ commutes with $H_{\mathrm{eff}}^{(k)}(t)$, $k=1,2,3$, it is easy to verify that $\langle\mu(t)|H_{\mathrm{eff}}^{(1)}(t)|\nu(t)\rangle=0$ for $t\in[0,\tau_{1}]$, $\langle\mu(t)|H_{\mathrm{eff}}^{(2)}(t)|\nu(t)\rangle=0$ for $t\in[\tau_{1},\tau_{2}]$, and $\langle\mu(t)|H_{\mathrm{eff}}^{(3)}(t)|\nu(t)\rangle=0$ for $t\in[\tau_{2},\tau]$, where $|\mu(t)\rangle=U(t,0)|\mu\rangle$, $|\nu(t)\rangle=U(t,0)|\nu\rangle$, and $|\mu\rangle,|\nu\rangle\in \mathcal{S}_3$. That is, condition $(\mathrm{ii})$ is satisfied too. Therefore, the unitary operator $U(\tau,0)$ plays a holonomic gate on the computational subspace.

{\it Third, we show how to realize a nonadiabatic holonomic $(n+1)$-qubit controlled-$(\boldsymbol{\mathrm{n}\cdot\sigma})$ gate $(n\geq3)$,}
\begin{align}\label{eq8}
U_{{C^n-\boldsymbol{\mathrm{n}\cdot\sigma}}}=\left[I^{\otimes n}-(|1\rangle\langle1|)^{\otimes n}\right]\otimes I+
(|1\rangle\langle1|)^{\otimes n}\otimes\boldsymbol{\mathrm{n}\cdot\sigma}.
\end{align}

To realize a nonadiabatic holonomic $(n+1)$-qubit controlled-$(\boldsymbol{\mathrm{n}\cdot\sigma})$, we consider $(n+1)$ three-level ions trapped in a line trap, and divide the whole evolution into $(2n-1)$ intervals: $0\le t \le \tau_1$, $\tau_1\le t \le \tau_2$,   $\cdots$,  $\tau_{2n-2}\le t \le \tau$.

In the first intervals, we take the effective Hamiltonian as,
\begin{align}
H_{\mathrm{eff}}^{(1)}(t)=\Omega_{11}(t)|ee\rangle\langle11|+\mathrm{H.c.},\label{T27}
\end{align}
which acts on ions $1$ and $2$.
In the $k$th interval $\tau_{k-1}\le t \le \tau_{k}$, $k=2,3,\cdots,n-1$,  we use the effective Hamiltonian,
\begin{align}
H_{\mathrm{eff}}^{(k)}(t)=\tilde{\Omega}_{e1}(t)|1e\rangle\langle e1|+\mathrm{H.c.},\label{T28}
\end{align}
which acts on ions $k$ and $k+1$.
In the $n$th interval $\tau_{n-1}\le t \le \tau_{n}$, we use the effective Hamiltonian,
\begin{align}
H_{\mathrm{eff}}^{(n)}(t)=\Omega_{e0}(t)|1e\rangle\langle e0|+\Omega_{e1}(t)|1e\rangle\langle e1|+\mathrm{H.c.},\label{T29}
\end{align}
which acts on ions $n$ and $n+1$.
In the $k$th interval $\tau_{k-1}\le t \le \tau_{k}$, $k=n+1, n+2,\cdots, 2n-2$, we use the effective Hamiltonian,
\begin{align}
H_{\mathrm{eff}}^{(k)}(t)=\tilde{\Omega}_{e1}(t)|1e\rangle\langle e1|+\mathrm{H.c.},\label{T30}
\end{align}
which acts on ions $(2n-k)$ and $(2n-k+1)$.
In the $(2n-1)$th interval $\tau_{2n-2}\le t \le \tau_{2n-1}$, we use the effective Hamiltonian,
\begin{align}
H_{\mathrm{eff}}^{(2n-1)}(t)=(-1)^{n}\Omega_{11}(t)|ee\rangle\langle11|+\mathrm{H.c.},\label{T31}
\end{align}
which acts on ions $1$ and $2$.

All these effective Hamiltonians can be realized by applying lasers on ions as those do in the two-qubit and three-qubit gates. In fact, the effective Hamiltonian in Eq. (\ref{T27}) is the same as the one in Eq. (\ref{T1}), the Hamiltonian in Eqs. (\ref{T28}) and (\ref{T30}) is a special case of that in Eq. (\ref{T2}), the Hamiltonian in Eq. (\ref{T29}) is the same as that in Eq. (\ref{T2}), and the Hamiltonian in Eq. (\ref{T31}) has the same form as that in Eq. (\ref{T1}) but only with a different parameter value.

To obtain a nonadiabatic holonomic $(n+1)$-qubit controlled-$(\boldsymbol{\mathrm{n}\cdot\sigma})$ gate with $\boldsymbol{\mathrm{n}} =(\sin\theta\cos\varphi,\sin\theta\sin\varphi,\cos\theta)$,
we set $\Omega_{11}(t)$ and $\tilde{\Omega}_{e1}(t)$ to be real, and set $\Omega_{e0}(t)=\Omega(t)\cos(\theta/2)$ and $\Omega_{e1}(t)=\Omega(t)\sin(\theta/2)e^{-i\varphi}$, where $\Omega(t)$ is a time dependent parameter. In this case, the evolution operator can be expressed as
\begin{align}
U(t,0)=U(t,\tau_{k-1})U(\tau_{k-1},\tau_{k-2})\cdots U(\tau_1,0),
\label{Tut}
\end{align}
for $t\in[\tau_{k-1},\tau_{k}]$, where $U(t,\tau_{a-1})=\exp\left[-i\int^{t}_{\tau_{a}}H_{\mathrm{eff}}^{(a)}(t^{\prime})dt^{\prime}\right]$ with $a=1,\cdots,2n-1$ and $\tau_0=0$.

We let  $|\mathrm{D_{n+1}}\rangle=\cos(\theta/2)|1^{\otimes n}0\rangle+\sin(\theta/2)e^{i\varphi}|1^{\otimes n}1\rangle$ and  $|\mathrm{B_{n+1}}\rangle=\sin(\theta/2)e^{-i\varphi}|1^{\otimes n}0\rangle-\cos(\theta/2)|1^{\otimes n}1\rangle$, which span the same subspace as $\textrm{Span}\{|1^{\otimes n}0\rangle,|1^{\otimes n}1\rangle\}$.  If we take $\tau_{1}$, $\tau_{2}$, $\cdots$, $\tau_{2n-2}$, and $\tau$ to satisfy
\begin{align}
&\int^{\tau_{1}}_{0}\Omega_{11}(t)dt=\int^{\tau}_{\tau_{2n-2}}\Omega_{11}(t)dt=\frac{\pi}{2},~~ \int^{\tau_{n}}_{\tau_{n-1}}\Omega(t)dt=\pi,\notag\\
&\int^{\tau_{k}}_{\tau_{k-1}}\tilde{\Omega}_{e1}(t)dt=\frac{\pi}{2},k=2,\cdots,n-1,n+1,\cdots, 2n-2,
\end{align}
there will be
\begin{align}
U(\tau,0)|\mathrm{D_{n+1}}\rangle=|\mathrm{D_{n+1}}\rangle,~~ U(\tau,0)|\mathrm{B_{n+1}}\rangle=-|\mathrm{B_{n+1}},\rangle,
\end{align}
and $U(\tau,0)|S\rangle=|S\rangle$ for $|S\rangle\in \{|b_1b_2\cdots b_{n+1}\rangle, ~b_i=0,1\}$ except $|1^{\otimes n}0\rangle$ and $|1^{\otimes n}1\rangle$.
If we focus only on the computational space,  the evolution operator $U(\tau,0)$ is equivalent to
\begin{align}
U_{{C^n-\boldsymbol{\mathrm{n}\cdot\sigma}}}=\left[I^{\otimes n}-(|1\rangle\langle1|)^{\otimes n}\right]\otimes I
+|\mathrm{D_{n+1}}\rangle\langle\mathrm{D_{n+1}})
-|\mathrm{B_{n+1}}\rangle\langle\mathrm{B_{{n+1}}}|.\label{ucn}
\end{align}
Since $|\mathrm{D_{n+1}}\rangle\langle\mathrm{D_{n+1}}|-
|\mathrm{B_{n+1}}\rangle\langle\mathrm{B_{n+1}}|=(|1\rangle\langle1|)^{\otimes n} \otimes\boldsymbol{\mathrm{n}\cdot\sigma}$, Eq. (\ref{ucn}) can be recast as Eq. (\ref{eq8}).

Similar to the two-qubit and three-qubit cases, we can demonstrate that $U_{{C^n-\boldsymbol{\mathrm{n}\cdot\sigma}}}$ is a holonomic gate. Indeed, the above discussion shows that a state initially residing in the computational space returns back to the computational space after the whole evolution. Furthermore, since $U(t,0)$  always commutes with the instantaneous effective Hamiltonian for each time interval, it is easy to verify that condition $\textrm{(ii)}$ is satisfied too. Therefore, $U_{{C^n-\boldsymbol{\mathrm{n}\cdot\sigma}}}$ is a holonomic gate.

So far, we have demonstrated how to realize a nonadiabatic holonomic $(n+1)$-qubit controlled-$(\boldsymbol{\mathrm{n}\cdot\sigma})$ gate. It is interesting to note that the well-known controlled-NOT gate and the Toffoli gate are the special cases of two-qubit and three-qubit controlled-$(\boldsymbol{\mathrm{n}\cdot\sigma})$ gates at $\boldsymbol{\mathrm{n}}=(1,0,0)$, respectively. Furthermore, we would like to point out that a nonadiabatic holonomic $(n+1)$-qubit controlled arbitrary rotation gate can be realized by combining two  controlled-$(\boldsymbol{\mathrm{n}\cdot\sigma})$ gates. Indeed, since $(\boldsymbol{\sigma\cdot}\boldsymbol{\mathrm{n}})(\boldsymbol{\sigma\cdot}\boldsymbol{\mathrm{m}})
=(\boldsymbol{\mathrm{n}\cdot \mathrm{m}})I
+i\boldsymbol{\sigma\cdot}(\boldsymbol{\mathrm{n}}\times\boldsymbol{\mathrm{m}})$,  we have
$U_{{C^n-\boldsymbol{\mathrm{n}\cdot\sigma}}}U_{{C^n-\boldsymbol{\mathrm{m}\cdot\sigma}}}=[I^{\otimes n}-(|1\rangle\langle1|)^{\otimes n}]\otimes I
+(|1\rangle\langle1|)^{\otimes n}\otimes[\boldsymbol{\mathrm{n}\cdot \mathrm{m}}
+i\boldsymbol{\sigma\cdot}(\boldsymbol{\mathrm{n}}\times\boldsymbol{\mathrm{m}})]\equiv U_{{C^n-[\boldsymbol{\mathrm{n}\cdot \mathrm{m}}
+i\boldsymbol{\sigma\cdot}(\boldsymbol{\mathrm{n}}\times\boldsymbol{\mathrm{m}})]}}$. Hence, by properly choosing $\boldsymbol{\mathrm{m}}$ and $\boldsymbol{\mathrm{n}}$, we can realize an arbitrary $(n+1)$-qubit controlled-$[(\boldsymbol{\mathrm{n}\cdot \mathrm{m}})I
+i\boldsymbol{\sigma\cdot}(\boldsymbol{\mathrm{n}}\times\boldsymbol{\mathrm{m}})]$ gate.

Before concluding, we would like to add a brief discussion on the feasibility of our scheme. Our scheme is based on the effective Hamiltonian $H_{\mathrm{eff}}(t)$, which is derived from the Hamiltonian $H(t)$ under the large detuning condition, and we have ignored the decay from $|e\rangle$ to $|0\rangle$ and $|1\rangle$. Hence, the large detuning approximation and the decay may affect the fidelity of the controlled gates. To illustrate the feasibility of our scheme, we calculate the fidelity $F=\langle\psi(\tau)|\rho(\tau)|\psi(\tau)\rangle$ for the controlled-NOT gate and the Toffoli gate, where $|\psi(\tau)\rangle=U_{{C^n-\boldsymbol{\mathrm{n}\cdot\sigma}}}|\psi(0)\rangle$ is obtained by directly using our controlled gates, and $\rho(\tau)$ is obtained by resolving the Lindblad equation with Hamiltonian $H(t)$ \cite{Tong2} and the Lindblad operators $L_{k}=\sqrt{\gamma_{e0}}|0\rangle\langle e|+\sqrt{\gamma_{e1}}|1\rangle\langle e|$.  Our physical model is trapped ions $^{40}\mathrm{Ca}^{+}$, of which the two Zeeman-split sublevels $4\mathrm{S}^{-1/2}_{1/2}$ and $4\mathrm{S}^{+1/2}_{1/2}$  are taken as qubit states $|0\rangle$ and $|1\rangle$ while $|e\rangle$ is encoded into $3\mathrm{D}_{5/2}$ \cite{Benhelm,Ballance}. The decay ratio of  $|e\rangle$ to $|0\rangle$ and $|1\rangle$ is taken as $\gamma_{e0}=\gamma_{e1}= 1/(2\tau_f)$ with the life time $\tau_f=1.2\mathrm{s}$ \cite{Barton2000}, and the Lamb-Dicke parameter and the additional detuning are taken as $\eta=0.044$ and  $\delta=2\pi\times50\mathrm{kHz}$ \cite{Benhelm}.  For the controlled-NOT gate, we choose the laser parameters $\Omega_{1}(t)=\Omega^{\prime}_{1}(t)=-\tilde{\Omega}_{0}(t)=-\tilde{\Omega}_{1}(t)=2\pi\times30\mathrm{KHz}$, which are experimentally achievable, and take $|\psi(0)\rangle=|10\rangle$. For  the Toffoli gate, we choose $\Omega_{1}(t)=\Omega^{\prime}_{1}(t)=-\tilde{\Omega}_{0}(t)=-\tilde{\Omega}_{1}(t)=2\pi\times30\mathrm{KHz}$ in the second time interval, and $\Omega^{\prime}_{1}(t)=\tilde{\Omega}_{1}=2\pi\times30\mathrm{KHz}$ in the first and third time intervals, and take $|\psi(0)\rangle=|110\rangle$. Our numerical result shows that the
fidelities are $99.17\%$ and $95.98\%$, respectively, for the controlled-NOT gate and Toffoli gate.

In conclusion, we have proposed an approach to realize nonadiabatic holonomic multiqubit controlled gates based on trapped ions by which a $(n+1)$-qubit controlled-$(\boldsymbol{\mathrm{n}\cdot\sigma})$ gate can be realized by $(2n-1)$ basic operations, whereas a nonadiabatic holonomic $(n+1)$-qubit controlled arbitrary rotation gate can be obtained by combining two such gates.
Comparing with the previous schemes of nonadiabatic holonomic computation in which a multiqubit controlled gate is built by using a large number of  universal elementary gates, our scheme greatly reduces the operations of nonadiabatic holonomic quantum computation.

\begin{acknowledgments}
P.Z.Z. acknowledges support from the National Natural Science Foundation of China through Grant No. 11575101. G.F.X. acknowledges support from the National Natural Science Foundation of China through Grant No. 11605104. D.M.T. acknowledges support from the National Natural Science Foundation of China though Grant No. 11775129 and the National Basic Research Program of China through Grant No. 2015CB921004.
\end{acknowledgments}


\begin{thebibliography}{99}
\bibitem{Sjoqvist} E. Sj\"oqvist, D. M. Tong, L. M. Andersson, B. Hessmo, M. Johansson, and K. Singh, New J. Phys. \textbf{14}, 103035 (2012).
\bibitem{Xu} G. F. Xu, J. Zhang, D. M. Tong, E. Sj\"oqvist, and L. C. Kwek, Phys. Rev. Lett. \textbf{109}, 170501 (2012).
\bibitem{Anandan} J. Anandan, Phys. Lett. A \textbf{133}, 171 (1988).
\bibitem{Jones} J. A. Jones, V. Vedral, A. Ekert, and G. Castagnoli, Nature (London) {\bf 403}, 869 (2000).
\bibitem{Berry} M. V. Berry, Proc. R. Soc. London, Ser. A {\bf 392}, 45 (1984).
\bibitem{Zanardi} P. Zanardi and M. Rasetti, Phys. Lett. A \textbf{264}, 94 (1999).
\bibitem{Duan} L. M. Duan, J. I. Cirac, and P. Zoller, Science \textbf{292}, 1695 (2001).
\bibitem{Wilczek} F. Wilczek and A. Zee, Phys. Rev. Lett. {\bf 52}, 2111 (1984).
\bibitem{WangXB} X. B. Wang and K. Matsumoto, Phys. Rev. Lett. {\bf 87}, 097901 (2001).
\bibitem{Zhu1}S. L. Zhu and Z. D. Wang, Phys. Rev. Lett. {\bf 89}, 097902 (2002).
\bibitem{Aharonov} Y. Aharonov and J. Anandan, Phys. Rev. Lett. {\bf 58}, 1593 (1987).
\bibitem{Xu2015} G. F. Xu, C. L. Liu, P. Z. Zhao, and D. M. Tong, Phys. Rev. A \textbf{92}, 052302 (2015).
\bibitem{E2016} E. Sj\"oqvist, Phys. Lett. A \textbf{380}, 65 (2016).
\bibitem{S2016} E. Herterich and E. Sj\"{o}qvist, Phys. Rev. A \textbf{94}, 052310 (2016).
\bibitem{XuGF2018} G. F. Xu, D. M. Tong, E. Sj\"oqvist, Phys. Rev. A \textbf{98}, 052315 (2018).
\bibitem{Johansson2012} M. Johansson, E. Sj\"{o}qvist, L. M. Andersson, M. Ericsson, B. Hessmo, K. Singh, and D. M. Tong, Phys. Rev. A \textbf{86}, 062322 (2012).
\bibitem{Spiegelberg2013} J. Spiegelberg and E. Sj\"{o}qvist, Phys. Rev. A \textbf{88}, 054301 (2013).
\bibitem{Liang} Z. T. Liang, Y. X. Du, W. Huang, Z. Y. Xue, and H. Yan, Phys. Rev. A \textbf{89}, 062312 (2014).
\bibitem{Zhang} J. Zhang, L. C. Kwek, E. Sj\"oqvist, D. M. Tong, and P. Zanardi, Phys. Rev. A \textbf{89}, 042302 (2014).
\bibitem{Mousolou2014} V. A. Mousolou, C. M. Canali, and E. Sj\"{o}qvist, New J. Phys. \textbf{16}, 013029 (2014).
\bibitem{ZhangT2} J. Zhang, T. H. Kyaw, D. M. Tong, E. Sj\"{o}qvist, and L. C. Kwek, Sci. Rep. {\bf 5}, 18414 (2015).
\bibitem{Xue} Z. Y. Xue, J. Zhou, and Z. D. Wang, Phys. Rev. A \textbf{92}, 022320 (2015).
\bibitem{You} Y. M. Wang, J. Zhang, C. F. Wu, J. Q. You, and G. Romero, Phys. Rev. A \textbf{94}, 012328 (2016).
\bibitem{Sun} C. F. Sun, G. C. Wang, C. F. Wu, H. D. Liu, X. L. Feng, J. L. Chen, and K. Xue, Sci. Rep. \textbf{6}, 20292 (2016).
\bibitem{Xue2016} Z. Y. Xue, J. Zhou, Y. M. Chu, and Y. Hu, Phys. Rev. A \textbf{94}, 022331 (2016).
\bibitem{Xue2017} Z. Y. Xue, F. L. Gu, Z. P. Hong, Z. H. Yang, D. W. Zhang, Y. Hu, and J. Q. You, Phys. Rev. Appl. \textbf{7}, 054022 (2017).
\bibitem{Zhao2017} P. Z. Zhao, G. F. Xu, Q. M. Ding, E. Sj\"{o}qvist, and D. M. Tong, Phys. Rev. A \textbf{95}, 062310 (2017).
\bibitem{Zhao} P. Z. Zhao, X. D. Cui, G. F. Xu, E. Sj\"{o}qvist, and D. M. Tong, Phys. Rev. A \textbf{96}, 052316 (2017).
\bibitem{Su2017} S. L. Su, Y. Z. Tian, H. Z. Shen, H. P Zang, E. J. Liang, and S. Zhang, Phys. Rev. A \textbf{96}, 042335 (2017).
\bibitem{Xu2017} G. F. Xu, P. Z. Zhao, T. H. Xing, E. Sj\"{o}qvist, and D. M. Tong, Phys. Rev. A \textbf{95}, 032311 (2017).
\bibitem{Xu2017PRA} G. F. Xu, P. Z. Zhao, D. M. Tong, and E. Sj\"{o}qvist, Phys. Rev. A \textbf{95}, 052349 (2017).
\bibitem{Mousolou2017} V. A. Mousolou, Phys. Rev. A \textbf{96}, 012307 (2017).
\bibitem{Zhao2018} P. Z. Zhao, X. Wu, T. H. Xing, G. F. Xu, and D. M. Tong, Phys. Rev. A \textbf{98}, 032313 (2018).
\bibitem{Xue2018} Z. P. Hong, B. J. Liu, J. Q. Cai, X. D. Zhang, Y. Hu, Z. D. Wang, and Z. Y. Xue, Phys. Rev. A \textbf{97}, 022332 (2018).
\bibitem{Zhang2018} J. Zhang, S. J. Devitt, J. Q. You, and F. Nori, Phys. Rev. A \textbf{97}, 022335 (2018).
\bibitem{Long} G. R. Feng, G. F. Xu, and G. L. Long, Phys. Rev. Lett. \textbf{110}, 190501 (2013).
\bibitem{Long2017} H. Li, Y. Liu, and G. L. Long, Sci. China-Phys. Mech. Astron. \textbf{60}, 080311 (2017).
\bibitem{Abdumalikov} A. A. Abdumalikov, J. M. Fink, K. Juliusson, M. Pechal, S. Berger, A. Wallraff, and S. Filipp, Nature (London) \textbf{496}, 482 (2013).
\bibitem{Xu2018} Y. Xu, W. Cai, Y. Ma, X. Mu, L. Hu, Tao Chen, H. Wang, Y. P. Song, Z. Y. Xue, Z. Q. Yin, and L. Sun, Phys. Rev. Lett. \textbf{121}, 110501 (2018).
\bibitem{Danilin} S. Danilin, A. Veps\"{a}l\"{a}inen, and G. S. Paraoanu, Phys. Scr. \textbf{93}, 055101 (2018).
\bibitem{Egger} D.J. Egger, M. Ganzhorn, G. Salis, A. Fuhrer, P. M\"{u}ller, P.Kl. Barkoutsos, N. Moll, I. Tavernelli, and S. Filipp, Phys. Rev. Appl. \textbf{11}, 014017 (2019).
\bibitem{Yan} T. X. Yan, B. J. Liu, K. Xu, C. Song, S. Liu, Z. S. Zhang, H. Deng, Z. G. Yan, H. Rong, K. Q. Huang, M. H. Yung, Y. Z. Chen, and D. P. Yu, Phys. Rev. Lett. \textbf{122}, 080501 (2019).
\bibitem{Yin} Z. X. Zhang, P. Z. Zhao, T. H. Wang, L. Xiang, Z. L. Jia, P. Duan, D. M. Tong, Y. Yin, G. P. Guo, arXiv:1811.06252.
\bibitem{Duan2014} C. Zu, W. B. Wang, L. He, W. G. Zhang, C. Y. Dai, F. Wang, and L. M. Duan, Nature (London) \textbf{514}, 72 (2014).
\bibitem{Arroyo} S. A. Camejo, A. Lazariev, S. W. Hell, and G. Balasubramanian, Nat. Commun. \textbf{5}, 4870 (2014).
\bibitem{Sekiguchi} Y. Sekiguchi, N. Niikura, R. Kuroiwa, H. Kano, and H. Kosaka, Nat. Photon. {\bf 11}, 309 (2017).
\bibitem{Zhou} Brian B. Zhou, Paul C. Jerger, V. O. Shkolnikov, F. J. Heremans, Guido Burkard, and David D. Awschalom, Phys. Rev. Lett. \textbf{119}, 140503 (2017).
\bibitem{Nagata} K. Nagata, K. Kuramitani, Y. Sekiguchi, and H. Kosaka, Nat. Commun. {\bf 9}, 3227 (2018).
\bibitem{Ishida} N. Ishida, T. Nakamura, T. Tanaka, S. Mishima, H. Kano, R. Kuroiwa, Y. Sekiguchi, and H. Kosaka, Opt. Lett. {\bf 43}, 2380 (2018).
\bibitem{Huangnew} Except for the elementary gates, only a three-qubit holonomic gate was given based on the shortcuts to adiabaticity in Ref. \cite{Huangnew2}.
\bibitem{Huangnew2}  B. H. Huang, Y. H. Kang, Z. C. Shi, J. Song, and Y. Xia, Ann. Phys. (Berlin) \textbf{530}, 1800179 (2018).
\bibitem{Shor} P. W. Shor, Phys. Rev. A \textbf{52}, 2493(R) (1995).
\bibitem{Steane} A. M. Steane, Phys. Rev. Lett. \textbf{77}, 793 (1996).
\bibitem{Grover} Lov K. Grover, Phys. Rev. Lett. \textbf{80}, 4329 (1998).
\bibitem{Vandersypen} L. M. K. Vandersypen, M. Steffen, G. Breyta, C. S. Yannoni, M. H. Sherwood, and I. L. Chuang, Nature (London) \textbf{414}, 883 (2001).
\bibitem{Joshi} A. Joshi and M. Xiao, Phys. Rev. A \textbf{74}, 052318 (2006).
\bibitem{Yang} W. L. Yang, C. Y. Chen, and M. Feng, Phys. Rev. A \textbf{76}, 054301 (2007).
\bibitem{Ota} Y. Ota, Y. Goto, Y. Kondo, and M. Nakahara, Phys. Rev. A \textbf{80}, 052311 (2009).
\bibitem{Barenco} A. Barenco, C. H. Bennett, R. Cleve, D. P. DiVincenzo, N. Margolus, P. Shor, T. Sleator, J. A. Smolin, and H. Weinfurter, Phys. Rev. A \textbf{52}, 3457 (1995).
\bibitem{Goto} H. Goto and K. Ichimura, Phys. Rev. A \textbf{70}, 012305 (2004).
\bibitem{SM1999} A. S{\o}rensen and K. M{\o}lmer, Phys. Rev. Lett. \textbf{82}, 1971 (1999).
\bibitem{SM2000} A. S{\o}rensen and K. M{\o}lmer, Phys. Rev. A \textbf{62}, 022311 (2000).
\bibitem{Benhelm} J. Benhelm, G. Kirchmair, C. F. Roos, and R. Blatt, Nat. Phys. \textbf{4}, 463 (2008).
\bibitem{Webb} A. E. Webb, S. C. Webster, S. Collingbourne, D. Bretaud, A. M. Lawrence, S. Weidt, F. Mintert, and W. K. Hensinger, Phys. Rev. Lett. \textbf{121}, 180501 (2018).
\bibitem{Shapira} Y. Shapira, R. Shaniv, T. Manovitz, N. Akerman, and R. Ozeri, Phys. Rev. Lett. \textbf{121}, 180502 (2018).
\bibitem{Tong1} By following the approach in Ref. \cite{James2007}, where it was proved that the effective Hamiltonian corresponding to the interaction one $H(t)=\sum_{n=1}^{N}(\hat{h}_n e^{-i\omega_nt}+\hat{h}^\dag_n e^{i\omega_nt})$ (i.e. Eq. (3.1) in Ref. \cite{James2007}) reads $H_{\mathrm{eff}}(t)=\sum_{m,n=1}^{N}\frac{1}{2\hbar} (\frac{1}{\omega_m}+\frac{1}{\omega_n})[\hat{h}^\dag_m,\hat{h}_n] e^{i(\omega_m-\omega_n)t}$ (i.e. Eq. (3.10) in Ref. \cite{James2007}), one can easily derive  Eq. (\ref{eq2})  from (\ref{eq1}) in our paper. Indeed, comparing Eq. (\ref{eq1}) in our paper with Eq. (3.1) in Ref. \cite{James2007}, we have
$\hat{h}_1=i\eta\Omega_{1}(t)a^{\dag}|e\rangle_{11}\langle1|$, $\hat{h}_2=i\eta\Omega^{\prime}_{1}(t)a|e\rangle_{11}\langle1|$,  $\hat{h}_3=
-i\eta\tilde{\Omega}_{0}^\ast(t)a^\dag|0\rangle_{22}\langle e|$, $\hat{h}_4=-i\eta\tilde{\Omega}_{1}^\ast(t)a|1\rangle_{22}\langle e|$, and $\omega_1=\omega_2=\omega_3=\omega_4=\delta$. Substituting them into (3.10), we immediately obtain  $H_{\mathrm{eff}}(t)=\Omega_{10}(t)|ee\rangle\langle10|+\Omega_{11}(t)|ee\rangle\langle11|+\mathrm{H.c}+\mathrm{Stark~ shift~terms}$. Noting that the Stark shift terms can be easily compensated by applying additional lasers \cite{Haffner2003}, we finally obtain the effective Hamiltonian in Eq.(\ref{eq2}).  Similarly, we can obtain the one in Eq. (\ref{T2}).
\bibitem{James2007} D. F. V. James and J. Jerke, Can. J. Phys. \textbf{85}, 625 (2007).
\bibitem{Haffner2003} H. H$\ddot{a}$ffner, S. Gulde, M. Riebe, G. Lancaster, C. Becher, J. Eschner, F. Schmidt-Kaler, and R. Blatt, Phys. Rev. Lett. 90, 143602 (2003).
\bibitem{Tong2} Note that additional lasers need to be introduced for compensating the Stark shift terms neglected in the effective Hamiltonians. For simplicity, we assume the vibrational mode is initially in the vacuum state in the numerical simulation.
\bibitem{Ballance} C. J. Ballance, V. M. Sch\"{a}fer, J. P. Home, D. J. Szwer, S. c. Webster, D. T. C. Allcock, N. M. Linke, T. P. Harty, D. P. L. A. Craik1, D. N. Stacey, A. M. Steane, and D. M. Lucas, Nature (London) \textbf{528}, 384 (2015).
\bibitem{Barton2000} P. A. Barton, C. J. S. Donald, D. M. Lucas, D. A. Stevens, A. M. Steane, and D. N. Stacey, Phys. Rev. A \textbf{62}, 032503 (2000).

\end{thebibliography}
\end{document}